%% file: ifacconf.tex
\documentclass{ifacconf}

\usepackage{graphicx}      
\usepackage{natbib}        
\usepackage{amssymb}
\usepackage{amsmath, bm}
\usepackage{mathrsfs}
\usepackage{tikz}
\usepackage[nolist]{acronym}
\usepackage{pgfplots}
\usepackage[ruled,vlined,titlenumbered,linesnumbered,algo2e]{algorithm2e}

\input{defs.tex}

\input{acronyms.tex}

\begin{document}
\begin{frontmatter}

\title{Fast Kötter--Nielsen--H\o holdt Interpolation over Skew Polynomial Rings\thanksref{footnoteinfo}} 

\thanks[footnoteinfo]{The authors would like to thank Johan Rosenkilde for valuable and fruitful discussions.}

\author[dlr]{Hannes Bartz}
\author[dlr]{Thomas Jerkovits} 

\address[dlr]{Institute of Communications and Navigation, \\
   German Aerospace Center (DLR), D-82234 Oberpfaffenhofen, Germany (e-mail: {\{hannes.bartz, thomas.jerkovits\}@dlr.de})}

\begin{abstract}                
   Skew polynomials are a class of non-commutative polynomials that have several applications in computer science, coding theory and cryptography.
   In particular, skew polynomials can be used to construct and decode evaluation codes in several metrics, like e.g. the Hamming, rank, sum-rank and skew metric.

   In this paper we propose a fast divide-and-conquer variant of the \acf{KNH} interpolation over free modules over skew polynomial rings.
   The proposed \ac{KNH} interpolation can be used to solve the interpolation step of interpolation-based decoding of (interleaved) Gabidulin, linearized Reed--Solomon and skew Reed--Solomon codes efficiently, which have various applications in coding theory and code-based quantum-resistant cryptography.
\end{abstract}

\begin{keyword}
   Kötter interpolation, skew polynomial rings, divide-and-conquer
\end{keyword}

\end{frontmatter}

\section{Introduction}
Skew polynomials are a class of non-commutative polynomials, that were introduced by~\cite{ore1933theory} and that have a variety of applications in computer science, coding theory and cryptography. 
The non-commutativity stems from the multiplication rule, which involves both, a field automorphism $\aut$ and a field derivation $\der$.
Unlike ordinary polynomials, there exist several ways to evaluate skew polynomials.
In~\cite{lam1985general,lam1988vandermonde} general results regarding the so-called \emph{remainder evaluation} of skew polynomials were defined. 
First results on the \emph{generalized operator evaluation} were derived in~\cite{leroy1995pseudolinear}.
Depending on the choice of the automorphism $\aut$ and the derivation $\der$, skew polynomial rings include several polynomial rings as a special case, such as the ordinary polynomial ring as well as the linearized polynomial ring~\cite{Ore_OnASpecialClassOfPolynomials_1933,ore1933theory}.
This property along with the different ways to evaluate skew polynomials make them a very versatile tool with many different applications.

One important application of skew polynomials is the construction of evaluation codes, that have distance properties in several decoding metrics, including the Hamming, rank, sum-rank, skew and other related metrics such as the (sum-)subspace metric~\cite{boucher2014linear,martinez2018skew,martinez2019reliable,caruso2019residues}.

In general, evaluation codes can be decoded via efficient interpolation-based decoding algorithms, like e.g. the~\cite{welch1986error} and~\cite{sudan1997decoding} algorithms for decoding Reed--Solomon codes.
In~\cite{koetter_dissertation} a bivariate interpolation algorithm for Sudan-like decoding of Reed--Solomon codes~\cite{koetter_dissertation} (over ordinary polynomial rings) was presented that since then is often referred to as the \emph{Kötter interpolation}.
The Kötter interpolation as it is known today was first stated by Nielsen and H{\o}holdt in~\cite{nielsen2000decoding} as a generalization of Kötter's algorithm~\cite{koetter_dissertation}.
To acknowledge the contribution by Nielsen and H{\o}holdt we refer to the algorithm as \acf{KNH} interpolation.
A fast divide-and-conquer variant of the~\ac{KNH} interpolation for the Guruswami--Sudan algorithm for decoding Reed--Solomon codes was presented in~\cite{nielsen2014fast}.

A multivariate generalization of the \ac{KNH} interpolation~\cite{wang2005kotter} for free modules over ordinary polynomial rings was proposed in~\cite{wang2005kotter}.
This approach was generalized to free modules over linearized polynomial rings by~\cite{xie2011general}. 
A further generalization to free modules over skew polynomial rings was proposed by~\cite{liu2014kotter}, which contains the variants over ordinary polynomial rings~\cite{wang2005kotter} and linearized polynomial rings~\cite{xie2011general} as special cases.

The evaluation and interpolation of multivariate skew polynomials was also considered in~\cite{martinez2019evaluation}; here with the main motivation to construct Reed-Muller-like codes (see also~\cite{geiselmann2019skew,martinez2020theory}).

In this paper, we propose a fast~\ac{DaC} variant of the \ac{KNH} interpolation over skew polynomial rings~\cite{liu2014kotter}, that uses ideas from~\cite{nielsen2014fast}.
The main idea of the proposed algorithm (Algorithm~\ref{alg:skewIntTree}) is, that the interpolation problem is divided into smaller sub-problems, that can be solved and merged efficiently.
In particular, the update operations in each loop of the \ac{KNH} interpolation are ``recorded'' and then applied to a degree-reduced basis in the merge step rather than to a non-reduced basis.
This allows to restrict the degree of the polynomials during the interpolation procedure which in turn results in a lower computational complexity.

We state the interpolation problem and the fast skew \ac{KNH} interpolation algorithm in a general way using linear functionals over skew polynomial rings with arbitrary automorphisms and derivations.
The proposed framework can be used for interpolation-based decoding of skew evaluation codes such as (linearized) Reed--Solomon and Gabidulin codes in several decoding metrics by an appropriate choice of the automorphism/derivation and definition of the linear functionals. The correctness proofs are omitted due to space restrictions and will be provided in a future full version of the paper.

The considered interpolation problem can also be solved using the skew minimal approximant bases methods from~\cite{bartz2020fast}.
Compared to the proposed approach, the approach from~\cite{bartz2020fast} requires the quite involved theory of minimal approximant bases in skew polynomial rings, whereas the proposed approach uses ideas from the well-known \ac{KNH} interpolation which is simpler and therefore easier to understand.

\section{Preliminaries}\label{sec:preliminaries}

Let $\Fq$ be a finite field and denote by $\Fqm$ the extension field of degree $m$. 
Let $\aut:\Fqm\mapsto\Fqm$ be a field automorphism of $\Fqm$ and let $\der:\Fqm\mapsto\Fqm$ be a $\aut$-derivation such that
\begin{align} 
   \der(a+b)\!=\!\der(a)\!+\!\der(b)
   \quad\text{and}\quad
   \der(ab)\!=\!\der(a)b\!+\!\aut(a)\der(b).
\end{align}

\emph{Skew polynomials} are non-commutative polynomials that were introduced by Ore~\cite{ore1933theory}. 
The set of all polynomials of the form 
\begin{equation}
f(x)=\sum_{i}f_ix^i \qquad \text{with }f_i\in\Fqm
\end{equation}
together with the ordinary polynomial addition and the multiplication rule
\begin{equation}\label{eq:mult_rule}
xa=\aut(a)x+\der(a)
\end{equation}
forms the non-commutative ring of \emph{skew polynomials} that is denoted by $\SkewPolyring$.
The \emph{degree} of a polynomial $f\in\SkewPolyring$ is defined as $\deg(f)\defeq\max_i\{i:f_i\neq0\}$ for $f\neq0$ and $-\infty$ else.
Further, by $\SkewPolyring_{<n}$ we denote the set of skew polynomials from $\SkewPolyring$ of degree less than $n$.

The \emph{monic}~\ac{LCLM} of some polynomials $p_0,p_1,\dots,p_{n-1}\in\SkewPolyring$ is denoted by
\begin{equation}\label{eq:def_lclm}
\lclm{p_i}{0\leq i\leq n-1}\defeq\lclm{p_0,p_1,\dots,p_{n-1}}{}.
\end{equation} 

The skew polynomial ring $\SkewPolyring$ is a left and right Euclidean domain.
Efficient Euclidean-like algorithms for performing left/right skew polynomial division exist~(see \cite{caruso2017fast,caruso2017new,puchinger2017fast}).
For two skew polynomials $f,g\in\SkewPolyring$, denote by $f\modr g$ the remainder of the right division of $f$ by $g$.

Skew polynomial rings include several polynomial rings as special cases, like e.g. ordinary polynomial rings and linearized polynomial rings.

The generalized operator evaluation defined in~\cite{leroy1995pseudolinear} allows to $\Fq$-linearize the skew polynomial evaluation and therefore establishes the link between the skew polynomial ring and the linearized polynomial ring (cf. ~\cite{Ore_OnASpecialClassOfPolynomials_1933,ore1933theory}).

\subsubsection{Skew Polynomial Vectors and Matrices}

For two vectors $\vec{a},\vec{b}\in\SkewPolyring^n$ we denote by $\lclm{\vec{a},\vec{b}}{}$ the element-wise \ac{LCLM} of the elements in $\a$ and $\b$ and by $\vec{a}\modr\vec{b}$ the element-wise right modulo operation, respectively.
For a vector $\vec{a}=\left(a_0,a_1,\dots,a_{n-1}\right)\in\SkewPolyring^n$ and a vector $\vec{w}=(w_0,w_1,\dots,w_{n-1})\in\mathbb{Z}_+^n$ we define its $\vec{w}$-weighted degree as
\begin{equation}\label{eq:def_w_deg} 
\deg_{\vec{w}}(\vec{a})\defeq\max_{0 \leq j \leq n-1}\{\deg(a_j)+w_j\}.
\end{equation}
Further, we define the $\vec{w}$-weighted monomial ordering $\wOrder$ on $\SkewPolyring^n$ such that we have
\begin{equation}\label{eq:top_order}
x^\ell\vec{e}_j\wOrder x^{\ell'}\vec{e}_{j'}
\end{equation}

if $\ell+w_j<\ell'+w_{j'}$ or if $\ell+w_j=\ell'+w_{j'}$ and $j<j'$, where $\vec{e}_j$ denotes the $j$-th unit vector over $\SkewPolyring$.
The definition of $\wOrder$ coincides with the $\vec{w}$-weighted \ac{TOP} ordering as defined in~\cite{adams1994introduction}.

For a vector $\vec{a} \in \SkewPolyring^{n} \setminus \{\0\}$ and weighting vector $\vec{w} = (w_0,\dots,w_{n-1}) \in \mathbb{Z}_+^n$, we define the $\vec{w}$-pivot index $\ind{\a}{\w}$ of $\vec{a}$ to be the largest index $j$ with $0 \leq j \leq n-1$ such that $\deg(a_j) + w_j = \deg_{\vec{w}}(\vec{a})$.
A matrix $\mat{A}\in\SkewPolyring^{a\times b}$ with $a \leq b$ is in (row) $\vec{w}$-ordered weak Popov form if the $\vec{w}$-pivot indices of its rows are strictly increasing in the row index.

A free $\SkewPolyring$-module is a module that has a basis consisting of $\SkewPolyring$-linearly independent elements. The rank of this module equals the cardinality of that basis.
In the following we consider particular bases for (left) $\SkewPolyring$-modules.

\begin{defn}
 Consider a \emph{left} $\SkewPolyring$-submodule $\module{M}$ of $\SkewPolyring^b$.
 For $\vec{w} \in \ZZ^a$, a left \ac{wowPB} is a full-rank matrix $\mat{A}\in\SkewPolyring^{a \times a}$ such that
 \begin{enumerate}
    \item $\mat{A}$ is in $\vec{w}$-ordered weak Popov form,
    \item the rows of $\mat{A}$ are a basis of $\module{M}$.
 \end{enumerate}
\end{defn}

We now consider the skew \ac{KNH} interpolation from~\cite{liu2014kotter}, which is the skew polynomial analogue of the \ac{KNH} interpolation over ordinary polynomial rings in~\cite{wang2005kotter}.
Note, that due to the isomorphism between $\SkewPolyring$ and the ring of linearized polynomials for $\aut$ being the Frobenius automorphism and $\der=0$ (zero derivations), the~\ac{KNH} variant over linearized polynomial rings in~\cite{xie2011general} can be seen as a special case of~\cite{liu2014kotter}.

Define the $\Fqm$-linear functionals $\vecEvNoInput{i}$
\begin{align}\label{eq:skew_eval_maps}
   \SkewPolyring^{\intParam+1}&\mapsto \Fqm \nonumber
   \\ 
   \vec{Q}&\mapsto\Fqm 
\end{align}
and the kernels
\begin{equation}
   \module{K}_i\defeq\{\vec{b}\in\SkewPolyring^{\intParam+1}:\vecEv{i}{\vec{b}}=0\}
\end{equation}
for all $i=0,\dots,n-1$.
For $0\leq i\leq n-1$ the intersection $\bar{\module{K}}_i\defeq\module{K}_0\cap\module{K}_1\cap\dots\cap\module{K}_i$ contains all vectors from $\SkewPolyring^{\intParam+1}$ that are mapped to zero under $\vecEvNoInput{0},\vecEvNoInput{1},\dots,\vecEvNoInput{i}$, i.e.
\begin{equation}
   \bar{\module{K}}_i=\{\vec{b}\in\SkewPolyring^{\intParam+1}:\vecEv{j}{\vec{b}}=0,\forall j=0,\dots,i\}.
\end{equation}

Under the assumption that the $\bar{\module{K}}_i$ are $\SkewPolyring$-submodules for all $i=0,\dots,n-1$ (see~\cite{liu2014kotter}) we can state the general skew polynomial vector interpolation problem.

\begin{prob}[General Vector Interpolation Problem]\label{prob:generalIntProblem}
   Giv\-en the integer $\intParam\in\mathbb{Z}_+$, a set of $\Fqm$-linear functionals $\set{E}=\{\vecEvNoInput{0},\dots,\vecEvNoInput{n-1}\}$ and a vector $\vec{w}\in\mathbb{Z}_+^{\intParam+1}$ compute a \ac{wowPB} for the left $\SkewPolyring$-module $\bar{\module{K}}_{n-1}$.
\end{prob}

Problem~\ref{prob:generalIntProblem} can be solved using a slightly modified variant of the multivariate skew \ac{KNH} interpolation from~\cite{liu2014kotter}.
Since the solution of Problem~\ref{prob:generalIntProblem} is a \ac{wowPB} for the interpolation module $\bar{\module{K}}_{n-1}$ instead of a single minimal polynomial vector, we modified the output of~\cite[Algorithm~1]{liu2014kotter} such that it returns a whole basis for the interpolation module $\bar{\module{K}}_{n-1}$. A similar approach was used in~\cite{bartz2017algebraic} to construct a basis for the interpolation module over linearized polynomial rings.

In the description of the algorithms we denote by $\text{TOP-}{\arg \min}\{d_j\}$ the $\arg\min$ operator that returns the smallest possible index $j$ to break ties, where $\{d_j\}$ is a set of integers. 
This property is needed to comply with the $\wOrder$-ordering described in~\eqref{eq:top_order}.
It is well-known that the Kötter interpolation computes a \ac{wowPB} (i.e. a minimal Gröbner basis w.r.t. the monomial ordering $\wOrder$) for the left $\SkewPolyring$-submodule $\intModule{n-1}$.
The modified multivariate skew \ac{KNH} interpolation is summarized in Algorithm~\ref{alg:skewMultVarKNH}.

\IncMargin{1.5em}
\begin{algorithm2e}
 \caption{Modified Skew \ac{KNH} Interpolation}\label{alg:skewMultVarKNH}
 \SetKwInOut{Input}{Input}\SetKwInOut{Output}{Output}
 \Input{
    A set $\{\vecEvNoInput{0},\vecEvNoInput{1},\dots,\vecEvNoInput{n-1}\}$ of vector evaluation maps \\ 
    A ``weighting'' vector $\vec{w}\in\posInt^{\intParam+1}$
 }
 \Output{A \ac{wowPB} $\mat{B}\in\SkewPolyring^{(\intParam+1)\times(\intParam+1)}$ for $\intModule{n-1}$
 }
 \textbf{Initialize:}
 $\mat{B}=\mat{I}_{\intParam+1}\in\SkewPolyring^{(\intParam+1)\times(\intParam+1)}$\\

 \BlankLine
 \For{$i\leftarrow 0$ \KwTo $n-1$}{
    \For{$j\leftarrow 0$ \KwTo $\intParam$}{
       $\Delta_{j}\gets \vecEv{i}{\vec{b}_j}$ \label{alg1:functional}
    }
    $\set{J}\gets \{j:\Delta_{j}\neq 0\}$ \\
    \If{$\set{J}\neq \emptyset$}{
       $j^{*}\gets \text{TOP-}\underset{j\in J}{\arg\min}\{\deg_{\vec{w}}(\vec{b}_{j})\}$ \\
       $\vec{b}^{*} \gets \vec{b}_{j^{*}}$ \\
       \For{$j\in \set{J}$}{
          \If{$j=j^{*}$}{
             $\vec{b}_{j} \gets \left(x-\frac{\vecEv{i}{x \vec{b}^{*}}}{\Delta_{j^{*}}}\right) \vec{b}^{*}$ \label{alg1:degreeinc} \tcc*[r]{degree-increasing step}               
          }
          \Else{   
             $\vec{b}_{j} \gets \vec{b}_{j} - \frac{\Delta_{j}}{\Delta_{j^{*}}}\vec{b}^{*}$ \label{alg1:crosseval} \tcc*[r]{cross-evaluation step}
          }
       }
    }
 }
 \Return{$\mat{B}$}
\end{algorithm2e}
\DecMargin{1.5em}

In each iteration of Algorithm~\ref{alg:skewMultVarKNH} (and so~\cite[Algorithm~1]{liu2014kotter}) there are three possible update steps:
\begin{enumerate}
 \item \emph{No update}: The vector $\b_j$ is not updated if $\b_j$ is in the kernel 
 \begin{equation*}
    \bar{\module{K}}_i=\{\vec{b}\in\SkewPolyring^{\intParam+1}:\vecEv{i}{\vec{b}}=0,\forall i=0,\dots,i\}
 \end{equation*}
 already, i.e. if $\Delta_j=\vecEv{i}{\vec{b}_j}=0$.
 
 \item \emph{Cross-evaluation} (or \emph{order-preserving}~\cite{liu2014kotter}) update: For any $\b_j$ that is not minimal w.r.t. $\wOrder$ (i.e. $j\neq j^*$) the cross-evaluation update (Line~\ref{alg1:crosseval}) is performed such that
 \begin{align*}
 \vecEvNoInput{i}\left(\vec{b}_{j} - \frac{\Delta_{j}}{\Delta_{j^{*}}}\vec{b}^{*}\right)
 =\vecEv{i}{\vec{b}_{j}} - \frac{\vecEv{i}{\vec{b}_{j}}}{\vecEv{i}{\vec{b}_{j^*}}}\vecEv{i}{\vec{b}_{j^*}}
 =0.
 \end{align*}
 Note, that the ($\w$-weighted) degree of $\b_j$ is \emph{not} increased by this update.
 
 \item \emph{Degree-increasing} (or \emph{order-increasing}~\cite{liu2014kotter}) update: For the minimal vector $\b_{j^*}\defeq \b^*$ w.r.t. $\wOrder$ the degree-increasing update (Line~\ref{alg1:degreeinc}) is performed such that
 \begin{align*}
 \vecEvNoInput{i}\left(\!\left(x\!-\!\frac{\vecEv{i}{x \vec{b}^{*}}}{\Delta_{j^{*}}}\right) \vec{b}^{*}\right)
 \!=\!\vecEv{i}{x\b^{*}}\!-\!\frac{\vecEv{i}{x \vec{b}^{*}}}{\vecEv{i}{\vec{b}^{*}}}\vecEv{i}{ \vec{b}^{*}}
 \!=\!0.
 \end{align*}
 The ($\w$-weighted) degree of $\b^*$ is increased by one in this case. 
\end{enumerate}
The different update steps are illustrated in~\cite[Figure~1]{liu2014kotter}.

\section{Fast Kötter--Nielsen--H\o holdt Interpolation over Skew Polynomial Rings}\label{sec:fast_skew_knh}

In~\cite{nielsen2014fast} a fast \ac{DaC} variant of the Kötter interpolation over ordinary polynomial rings for the Guruswami--Sudan decoder was presented.
We now use ideas from~\cite{nielsen2014fast} to speed up the skew \ac{KNH} interpolation from~\cite{liu2014kotter}. 
The main idea is to split up Problem~\ref{prob:generalIntProblem} into smaller subproblems and to consider degree-reduced skew polynomial matrices rather than the full matrices as in~\cite{liu2014kotter}.
In the following, we describe the general framework for the fast skew~\ac{KNH} interpolation algorithm w.r.t. to general vector evaluation maps based on e.g. the generalized operator evaluation and remainder evaluation.

The operations performed on the basis $\mat{B}$ in the inner loop of the $i$-th iteration of Algorithm~\ref{alg:skewMultVarKNH} can be represented by the matrix
\begin{equation}\label{eq:opMatrix}
\arraycolsep=4pt
\mat{U}_i=
\left(
\begin{array}{ccc|c|ccc}
1 & & & -\frac{\Delta_0}{\Delta_{j^{*}}} & & &
\\[-3pt]
& \ddots & & \vdots & & &
\\
& & 1 & -\frac{\Delta_{j^{*}-1}}{\Delta_{j^{*}}} & & &
\\
& &   & x-\frac{\vecEv{i}{x \vec{b}_{j^{*}}}}{\Delta_{j^{*}}} & & &
\\
& & & -\frac{\Delta_{j^{*}+1}}{\Delta_{j^{*}}} & 1 & &
\\[-3pt]
& & & \vdots &  & \ddots &
\\
& & & -\frac{\Delta_{\intParam}}{\Delta_{j^{*}}} &  & & 1
\end{array}
\right)
\end{equation}
Note, that if $\Delta_j=0$ no update on the row $\vec{b}_j$ is performed since $\Delta_j/\Delta_{j^{*}}=0$. 

To describe the following algorithms we introduce some notations. $\vec{M}_{[i,j]} \in \SkewPolyring^{\intParam+1}$ denotes a polynomial vector that is dependent on the index set $\{i,i+1,\ldots,j-1,j\}$ with $j\geq i$ and $\vec{M}_{[i,i]} = \vec{M}_i$.
The set $\mathcal{M}$ is globally available for all algorithms and is defined as
\begin{align}
\set{M}=\{&\vec{M}_{[0,n-1]},\vec{M}_{[0,\lfloor n/2\rfloor-1]},\vec{M}_{[\lfloor n/2\rfloor,n-1]},\ldots \notag
\\
&\ldots,\vec{M}_{0},\vec{M}_{1},\ldots,\vec{M}_{n-1}\}\subseteq\SkewPolyring^{\intParam+1}\label{eq:min_poly_vec_set}
\end{align}
for an integer $n\in\posInt$.
For a set of evaluation maps $\set{E} = \{\vecEvNoInput{0},\ldots,\vecEvNoInput{n-1}\}$ we use a similar notation to access a subset of $\set{E}$ as $\set{E}_{[i,j]}=\{\vecEvNoInput{i},\dots,\vecEvNoInput{j}\}$.

In order to describe a general framework for the fast skew~\ac{KNH} interpolation, we need the following assumption for the linear functionals. 

\begin{assumption}\label{ass:mod_vectors}
Let $\set{E}=\{\vecEvNoInput{0},\ldots,\vecEvNoInput{n-1}\}$ be a set of linear functionals as defined in~\eqref{eq:skew_eval_maps} and let $\set{E}_{[i,j]}=\{\vecEvNoInput{i},\ldots,\vecEvNoInput{j}\}\subseteq\set{E}$. We assume that for all $0\leq i\leq j\leq n-1$ there exits a skew polynomial vector $\vecMinpolyNoX{[i,j]}\in\set{M}$ (which contains minimal skew polynomials that depend on $\set{E}_{[i,j]}$) such that
 \begin{equation}
  \vecEv{l}{\vec{Q}}=\vecEv{l}{\vec{Q}\modr\vecMinpolyNoX{[i,j]}},
  \quad\forall l=i,\dots,j.
 \end{equation}
\end{assumption}

\IncMargin{1.5em}
\begin{algorithm2e}
   \caption{\textsf{SkewInterpolatePoint}}\label{alg:skewIntPoint}
   \SetKwInOut{Input}{Input}\SetKwInOut{Output}{Output}
    \Input{A skew vector evaluation map $\vecEvNoInput{i}$,\\ 
    $\mat{B} \in \SkewPolyring^{(\intParam+1)\times(\intParam+1)}$\\
    $\vec{d}\in \posInt^{\intParam+1}$ \\ s.t. $d_{j}=\deg_{\vec{w}}(\vec{b}_j), \, \forall j=0,\dots,\intParam$.
    }
 \BlankLine
 \Output{$\mat{T}\in\SkewPolyring^{(\intParam+1)\times(\intParam+1)}$ s.t. the rows of $\mat{\hat{B}}\defeq\mat{TB}$ is a \ac{wowPB} for $\spannedBy{\mat{B}}\cap\module{K}_i$,\\ 
 $\vec{\hat{d}}\in \posInt^{\intParam+1}$ \\ s.t. $\hat{d}_j=\deg_{\vec{w}}(\vec{\hat{b}}_j),\,\forall j=0,\dots,\intParam$.
 }

   \BlankLine
   $\vec{\hat{d}} \gets \vec{{d}}$ \\
   \For{$j\leftarrow 0$ \KwTo $\intParam$}{
      $\Delta_{j}\gets \vecEv{i}{\vec{b}_j}$
   }
   $\set{J}\gets \{j:d_{j}\neq 0\}$ \\
   $\mat{T}\gets\mat{I}_{\intParam+1}\in\SkewPolyring^{(\intParam+1)\times(\intParam+1)}$ \\
   \If{$\set{J}\neq \emptyset$}{
         $j^{*}\gets\text{TOP-}\underset{l\in J}{\arg \min}\{d_l\}$ \\
         $\mat{T}\gets\mat{U}_i$ where $\mat{U}_i$ is as in~\eqref{eq:opMatrix}\\
         $\hat{d}_{j^{*}}\gets\hat{d}_{j^{*}}+1$
   }
   \Return{$(\mat{T}, \vec{\hat{d}})$}
\end{algorithm2e}
\DecMargin{1.5em}

Equipped with the routine \textsf{SkewInterpolatePoint} in Algorithm~\ref{alg:skewIntPoint} to solve the basic step we can now state a~\ac{DaC} variant of the skew~\ac{KNH} interpolation from~\cite{liu2014kotter}, which is given in Algorithm~\ref{alg:skewIntTree}.

\IncMargin{1.5em}
\begin{algorithm2e}
 \caption{\textsf{SkewInterpolateTree}}\label{alg:skewIntTree}
 \SetKwInOut{Input}{Input}\SetKwInOut{Output}{Output}
 \Input{linear functionals $\set{E}_{[i_1,i_2]}=\{\vecEvNoInput{i_1},\dots,\vecEvNoInput{i_2}\}$ \\
 $\mat{B}\in\SkewPolyring^{(\intParam+1)\times(\intParam+1)}$, \\
 $\vec{d}\in \posInt^{\intParam+1}$\\ s.t. $d_j=\deg_{\vec{w}}(\vec{b}_j), \, \forall j=0,\dots,\intParam$.
 }
 \BlankLine

 \Output{
 A matrix $\mat{T}\in\SkewPolyring^{(\intParam+1)\times(\intParam+1)}$ \\s.t. $\mat{\hat{B}}\!\defeq\!\mat{TB}$ is a \ac{wowPB} \\ for $\spannedBy{\mat{B}}\!\cap\!\module{K}_{i_1}\!\!\cap\!\dots\!\cap\!\module{K}_{i_2}$, \\ 
 $\vec{\hat{d}} \in \posInt^{\intParam+1}$ \\ s.t. $\hat{d}_j=\deg_{\vec{w}}(\vec{\hat{b}}_j),\,\forall j=0,\dots,\intParam$.
 }

 \BlankLine
 \If{$i_1=i_2$}{
    \Return{$\textsf{SkewInterpolatePoint}(\vecEvNoInput{i_1}, {\mat{B}}, \vec{d})$}
 }
 \Else{
    $z\gets\left\lfloor\frac{i_1+i_2}{2}\right\rfloor$ \\
    ${\mat{B}}_1\gets{\mat{B}}\modr\vecMinpolyNoX{[i_1,z]}$ \label{line:VecMod1} \\%
    $(\mat{T}_1, \vec{d}_1)\gets \textsf{SkewInterpolateTree}(\set{E}_{[i_1,z]}, {\mat{B}}_1, \vec{d})$ \\
    ${\mat{B}}_2\gets\mat{T}_1{\mat{B}}\modr \vecMinpolyNoX{[z+1,i_2]}$ \label{line:VecMod2} \\
    $(\mat{T}_2, \vec{d}_2)\gets \textsf{SkewInterpolateTree}(\set{E}_{[z+1,i_2]}, {\mat{B}}_2, \vec{d}_1)$ \\ 
    \Return{$(\mat{T}=\mat{T}_2\mat{T}_1, \vec{\hat{d}}=\vec{d}_2)$}\label{line:T2T1}
 }
\end{algorithm2e}
\DecMargin{1.5em}

An essential part in Algorithm~\ref{alg:skewIntTree} is the degree-reduction via the minimal skew polynomial vectors $\vecMinpolyNoX{[i,j]}$ from the globally available set $\set{M}$ defined in~\eqref{eq:min_poly_vec_set}.
We now sketch a generic procedure to pre-compute the set $\set{M}$ efficiently.
We consider minimal polynomials which can be constructed by means of the \ac{LCLM} of polynomial sequences, such as minimal skew polynomials with respect to the generalized operator and remainder evaluation (see~\cite[Theorem~3.2.7]{caruso2017new}).

\begin{figure}[ht!]
 \centering
 \input{tikz/vec_msp_tree.tikz}
 \caption{Illustration of the computation tree to precompute all minimal polynomial vectors in the set $\mathcal{M}$ for $\kappa=4$.}
 \label{fig:minpoly_vec_tree}
\end{figure}
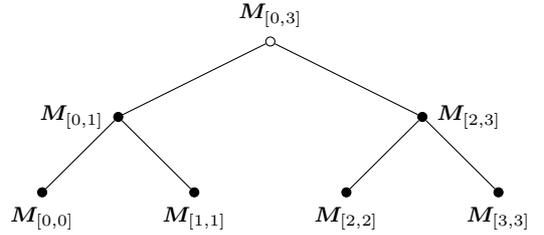

The \ac{DaC} structure of the proposed procedure is illustrated in Figure~\ref{fig:minpoly_vec_tree} for an example of $\kappa=4$. 
The initial minimal polynomial vectors $\vecMinpolyNoX{0},\vecMinpolyNoX{1},\ldots,\vecMinpolyNoX{\kappa-1}$ from which all other minimal polynomials are computed via the \ac{LCLM}, are computed w.r.t. to the generalized operator or remainder evaluation depending on the application. 

\vspace*{-10pt}

\section{Applications in Coding and Cryptography}\label{sec:applications}

Efficient methods for solving instances of Problem~\ref{prob:generalIntProblem} are essential for several applications in coding theory and code-based quantum-resistant cryptography.

For $\der=0$, instances of Problem~\ref{prob:generalIntProblem} correspond to the com\-plex\-i\-ty-dominating inter\-polation problem of interpolation-based decoding of (interleaved/folded) linearized Reed--Solomon codes in the sum-rank metric (see~\cite{martinez2018skew,caruso2019residues,bartz2021fast,hormann2021efficient}). 
Compared to the original skew \ac{KNH} interpolation by~\cite{liu2014kotter}, which requires at most $\OCompl{\intParam^2n^2}$ operations in $\Fqm$, Algorithm~\ref{alg:skewIntTree} can solve the corresponding interpolation problem requiring only at most $\softoh{\intOrder^\matmult\OMul{n}}$ operations in $\Fqm$, where $\intOrder\ll n$ is the interleaving order (or the decoding parameter for folded codes), $\matmult$ the matrix multiplication exponent (currently $\omega < 2.37286$), $n$ the length of the code and $\softoh{\cdot}$ the soft-$O$ notation which neglects log factors.

As (interleaved/folded) linearized Reed--Solomon codes include (interleaved/folded) Reed--Solomon codes (see~\cite{bleichenbacher2003decoding,guruswami2008explicit}) in the Hamming metric and (interleaved/folded) Gabidulin codes (see~\cite{overbeck2006decoding,mahdavifar2012list}) in the rank metric as special cases, the interpolation steps in the respective interpolation-based decoders (see e.g.~\cite{guruswami2008explicit,wachter2014list,mahdavifar2012list}) can be sped-up by Algorithm~\ref{alg:skewIntTree} to at most $\softoh{\intOrder^\matmult\OMul{n}}$ operations in $\Fqm$.

Similarly, the interpolation step in the interpolation-based decoder for ($\intOrder$-interleaved) skew Reed--Solomon codes in the skew metric can be solved using Algorithm~\ref{alg:skewIntTree} requiring at most $\softoh{\intOrder^\matmult\OMul{n}}$ operations in $\Fqm$.

Apart from error correction in communication systems, variants of (linearized) Reed--Solomon and Gabidulin codes are used to design quantum-resistant code-based cryptosystems (see e.g.~\cite{melchor2018hamming,melchor2017rank,Loidreau2017-NewRankMetricBased,renner2021liga,puchinger2020generic}).
In order to obtain a good encryption/decryption performance of the corresponding cryptosystems, fast decoding algorithms for the respective codes are required.
As in general the interpolation step dominates the overall complexity of interpolation-based decoders, the proposed fast skew \ac{KNH} interpolation algorithm is an important factor for developing high-performance code-based cryptosystems.

In order to prevent side-channel attacks that exploit correlations between the error patterns and the runtime of the interpolation algorithm, future work will consider constant-time variants of the proposed algorithm in the spirit of~\cite{bettaieb2019preventing}.

\section{Conclusion}\label{sec:conclusion}

We presented a fast divide-and-conquer variant of the \acf{KNH} interpolation over free modules over skew polynomial rings.
The proposed algorithm divides the interpolation problem into smaller sub-problems, which can be solved and merged efficiently, and uses degree-restricted polynomials for the intermediate steps to reduce the computational complexity.
The fast interpolation algorithm can be used to speed up existing interpolation-based decoders for variants of linearized and skew Reed--Solomon codes in the Hamming, (sum-)rank and skew metric.


\input{refs.bbl}
\end{document}

%% file: defs.tex
\newtheorem{assumption}{Assumption}

\pgfplotsset{compat=1.17}

\newcommand{\Fqm}{\ensuremath{\mathbb F_{q^m}}}

\newcommand{\Fq}{\ensuremath{\mathbb F_{q}}}

\newcommand{\ZZ}{\ensuremath{\mathbb{Z}}}
\newcommand{\set}[1]{\ensuremath{\mathcal{#1}}}

\newcommand{\spannedBy}[1]{\ensuremath{\langle #1\rangle}}

\newcommand{\aut}{\ensuremath{\sigma}}
\newcommand{\der}{\ensuremath{\delta}}
\newcommand{\SkewPolyring}{\ensuremath{\Fqm[x;\aut,\der]}}

\newcommand{\OCompl}[1]{\ensuremath{\mathcal{O}({#1})}}

\DeclareMathOperator{\defi}{def}
\newcommand{\defeq}{\overset{\defi}{=}}

\renewcommand{\bar}{\overline}

\newcommand{\modr}{\; \mathrm{mod}_\mathrm{r} \;}

\newcommand{\ind}[2]{\ensuremath{\textrm{Ind}_{\vec{#2}}(#1)}}

\renewcommand{\vec}[1]{\ensuremath{\mathbf{#1}}}
\newcommand{\mat}[1]{\ensuremath{\mathbf{#1}}}

\newcommand{\lclm}[2]{\ensuremath{\text{lclm}\left(#1\right)_{#2}}}

\renewcommand{\a}{\mathbf a}
\renewcommand{\b}{\mathbf b}

\newcommand{\s}{\mathbf s}

\newcommand{\w}{\mathbf w}

\renewcommand{\S}{\mathbf S}

\newcommand{\0}{\mathbf 0}

\newcommand{\vecMinpolyNoX}[1]{\ensuremath{{\vec{M}}_{#1}}}

\newcommand{\bnd}[2]{\ensuremath{#1\mathopen{}\left(#2\right)\mathclose{}}}

\newcommand{\softoh}[1]{\bnd{\widetilde{O}}{#1}}
\newcommand{\matmult}{\ensuremath{\omega}}
\newcommand{\OMul}[1]{\mathfrak{M}(#1)}

\newcommand{\intOrder}{\ensuremath{s}}

\newcommand{\posInt}{\ensuremath{\mathbb{Z}_{+}}} 

\newcommand{\vecEv}[2]{\ensuremath{\mathscr{E}_{#1}(#2)}}
\newcommand{\vecEvNoInput}[1]{\ensuremath{\mathscr{E}_{#1}}}

\newcommand{\intParam}{\ensuremath{s}}
\newcommand{\wOrder}{\ensuremath{\prec_{\vec{w}}}}
\newcommand{\module}[1]{\ensuremath{\set{#1}}}
\newcommand{\intModule}[1]{\ensuremath{\bar{\module{K}}_{#1}}}

%% file: acronyms.tex
\begin{acronym}
 \acro{TOP}{term-over-position}
 \acro{DaC}{divide-and-conquer}
 \acro{KNH}{Kötter--Nielsen--H\o holdt}
 \acro{LCLM}{least common left multiple}
 \acro{SRS}{skew Reed--Solomon}
 \acro{ISRS}{interleaved skew Reed--Solomon}
 \acro{LRS}{linearized Reed--Solomon}
 \acro{LLRS}{lifted linearized Reed--Solomon}
 \acro{ILRS}{interleaved linearized Reed--Solomon}
 \acro{LILRS}{lifted interleaved linearized Reed--Solomon}
 \acro{MSRD}{maximum sum-rank distance}
 \acro{MSD}{maximum skew distance}
 \acro{MRD}{maximum rank distance}
 \acro{wowPB}[{$\w$}-owPB]{$\vec{w}$-ordered weak-Popov Basis}
\end{acronym}

%% file: tikz/vec_msp_tree.tikz
\tikzset{
  solid node/.style={circle,draw,inner sep=1.2,fill=black},
  hollow node/.style={circle,draw,inner sep=1.2},
}

\begin{tikzpicture}[font=\footnotesize]
  \tikzset{
    level 1/.style={level distance=10mm,sibling distance=40mm},
    level 2/.style={level distance=10mm,sibling distance=20mm},
    level 3/.style={level distance=10mm,sibling distance=10mm},
    level 4/.style={level distance=10mm,sibling distance=5mm},
  }

  \node[hollow node,label=above:{$\vecMinpolyNoX{[0,3]}{}$}] (root) {}
    child{node[solid node,label=left:{$\vecMinpolyNoX{[0,1]}{}$}]{}
      child{node(l1)[solid node,label=below:{$\vecMinpolyNoX{[0,0]}$},label=below:{}]{}
        edge from parent node[left]{}
      }
      child{node(l2)[solid node,label=below:{$\vecMinpolyNoX{[1,1]}$},label=below:{}]{}
        edge from parent node[right]{}
      }
      edge from parent node[left,xshift=-10]{}
    }
    child{node[solid node,label=right:{$\vecMinpolyNoX{[2,3]}{}$}]{}
      child{node(r1)[solid node,label=below:{$\vecMinpolyNoX{[2,2]}$}]{}
        edge from parent node[left]{}
      }
      child{node(r2)[solid node,label=below:{$\vecMinpolyNoX{[3,3]}$}]{}
        edge from parent node[right]{}
      }
      edge from parent node[right,xshift=10]{}
    }
  ;
  \node[above of = root]{};

\end{tikzpicture}